\documentclass[USenglish,twocolumn]{article}
\PassOptionsToPackage{hyphens}{url}
\PassOptionsToPackage{bottom}{footmisc}
\usepackage[utf8]{inputenc}
\usepackage[T1]{fontenc}
\usepackage[USenglish]{babel}
\usepackage[sort&compress,square,numbers]{natbib}
\usepackage{times}
\usepackage{booktabs}
\usepackage{xcolor}
\usepackage{authblk}
\usepackage{abstract}
\usepackage{amsmath,amssymb}
\usepackage{url}
\usepackage[section]{placeins}

\begin{document}

\pagenumbering{gobble}

\title{Comparison of Dimension Reduction Methods for EEG Seizure Detection Using Autonomous AI-Driven Optimization}
\author[1]{Annika Stiehl}
\author[2]{Vishal Kagade}
\author[2]{Nicolas Weeger}
\author[3]{Nicole Ille}
\author[2]{Stefan Geißelsöder}
\author[2]{Christian Uhl}

\affil[1]{Ansbach University of Applied Sciences, Residenzstr. 8, Ansbach, Germany; annika.stiehl@hs-ansbach.de}
\affil[2]{Ansbach University of Applied Sciences, Ansbach, Germany}
\affil[3]{BESA GmbH, Gräfelfing, Germany}

\date{}

\maketitle

\begin{abstract}
  Automated epileptic seizure detection from multichannel electroencephalography (EEG) benefits from dimension reduction to obtain compact, discriminative representations. We compare four signal-space dimension reduction methods, Principal Component Analysis (PCA), Dynamical Component Analysis (DyCA), Dynamic Mode Decomposition (DMD), and Average Volatility Dimensioning (AVD), for deep learning-based seizure detection on the Temple University Hospital Seizure Corpus (TUSZ~v2.0.3). To enable a comparison of optimal combinations of representation and classifier, an autonomous AI-driven research framework independently optimizes architecture and hyperparameters for each representation. Measured by test ROC-AUC, the variance-based methods AVD (88.28\,\%) and PCA (85.98\,\%) paired with their respective optimal classifiers outperform the dynamics-based methods DMD (74.56\,\%) and DyCA (74.85\,\%) by over 10\,\%, with AVD also showing the smallest validation-to-test gap. The best-performing classifier architecture differs across representations, indicating that representation and classifier should be optimized jointly. Our results highlight the importance of the input representation for EEG seizure detection and indicate the viability of autonomous AI-driven experimentation in biomedical signal processing. 
\end{abstract}


\section{Introduction}

Long-term electroencephalography (EEG) monitoring generates more data than clinicians can manually review, making automated seizure detection essential for scalable clinical diagnostics and treatment~\cite{Shoeb2009}. While most research focuses on classifier architectures, the upstream choice of input representation is equally critical but rarely examined systematically, since any discriminative information discarded at this stage cannot be recovered by a downstream classifier, regardless of its capacity. One paradigm is feature-space reduction, where handcrafted features are extracted and then projected to lower dimensions~\cite{Anuragi2024,Zubair2021,Ventura2009}. An alternative is signal-space dimension reduction, which transforms the EEG signals themselves to a more expressive subspace. Several such methods exist with fundamentally different assumptions: Principal Component Analysis (PCA) preserves maximum variance~\cite{Jolliffe2010}; Dynamical Component Analysis (DyCA)~\cite{Uhl2024} and Dynamic Mode Decomposition (DMD)~\cite{Kutz2016} preserve temporal dynamics based on ODE or Koopman operator models, respectively; and Average Volatility Dimensioning (AVD) captures variance structure robust to outliers~\cite{Mallinger2026}. DyCA-based representations have been shown to capture seizure-relevant dynamics~\cite{Stiehl2023,Laemmermann2025}, and the noise robustness of DyCA against DMD has been systematically characterized on simulated attractors~\cite{Stiehl2025}. Yet how these dynamics-based descriptors compare against variance-based alternatives when paired with independently optimized classifiers has not yet been established.

We address this gap on the TUSZ corpus~v2.0.3 using an autonomous artificial intelligence (AI)-driven research framework that independently optimizes the classifier for each representation. Our contributions are: (1) a systematic comparison of PCA, DyCA, DMD, and AVD for seizure detection with independently optimized deep learning classifiers; and (2) an adaptation of an existing autonomous research framework to EEG seizure detection, showing its practical viability for hypothesis-driven model design on biomedical signal data.

\section{Methods} 

\subsection{Data and Preprocessing}

Experiments were conducted using the Temple University Hospital Seizure Corpus (TUSZ~v2.0.3)~\cite{Shah2018}, a publicly available repository of expert-annotated clinical scalp EEG recordings. Our analysis was restricted to recordings from adult subjects with a minimum duration of 3\,s, using the predefined, subject-disjoint training, validation, and test splits, so that no subject contributes to more than one split. After this restriction, the training, validation, and test splits comprised 547/48/42 subjects (1135/328/122 recording sessions), respectively.

The raw EEG signals were resampled to 256\,Hz and subjected to a filtering pipeline: a second-order infinite impulse response (IIR) notch filter at 60\,Hz removed line noise, while a first-order forward Butterworth highpass filter (1\,Hz, 6\,dB/oct) and a fourth-order zero-phase Butterworth lowpass filter (25\,Hz, 24\,dB/oct) defined the frequency band of interest. To ensure spatial consistency, all signals were remontaged to a 27-channel virtual average montage using the software package BESA Research.

For model input, continuous recordings were segmented into 2-second windows (512~samples). For the training set, these windows were extracted with a 0.5\,s overlap for data augmentation, while for the validation and test sets, non-overlapping windows were used. Each channel underwent robust scaling, determined by the 5th to 95th percentile range, and was subsequently clipped to the interval $[-0.5,\,0.5]$. For binary classification, a window was labeled as seizure if any part of it overlapped with an expert-annotated seizure interval, and as non-seizure otherwise. This yielded 1\,409\,505/94\,156/233\,397 windows for the training, validation, and test split, respectively. The training set is highly imbalanced, with only 7.9\,\% of windows labeled as seizure. As we adopt the predefined subject-disjoint splits without rebalancing, the seizure density is not matched across splits, differing markedly between training, validation, and test (7.9/33.7/5.4\,\%). These preprocessed 27-channel windows served as input to the dimension reduction methods, which projected them to lower-dimensional representations before classification.

\subsection{Dimension Reduction}

All four methods reduce the 27-channel EEG to $n\!=\!4$ input channels for the classifier, a choice informed by preliminary experiments that showed no benefit from higher dimensionalities. Transforms are fitted for each 2-second window independently. For PCA, DMD, and DyCA, the reduction takes the common form
\begin{equation}
\mathbf{Q} = \mathbf{M}\mathbf{X} + \mathbf{R},
\end{equation}
where $\mathbf{Q}\in\mathbb{R}^{N\times T}$ ($N\!=\!27$) is the multichannel window, $\mathbf{M}\in\mathbb{R}^{N\times n}$ collects $n$ spatial modes, $\mathbf{X}\in\mathbb{R}^{n\times T}$ is the low-dimensional signal, and $\mathbf{R}$ is the discarded residual. The methods differ in what $\mathbf{M}$ optimizes:

\textbf{PCA} selects the $n$ eigenvectors of the covariance matrix with the largest eigenvalues, thereby maximizing preserved variance~\cite{Jolliffe2010}.

\textbf{DMD} identifies modes of the best-fit linear operator $\mathbf{A}$ satisfying $\mathbf{Q}'\!\approx\!\mathbf{A}\mathbf{Q}$ (consecutive time steps), isolating frequency-resolved dynamic structures rather than variance-dominant ones~\cite{Kutz2016}. Complex-conjugate mode pairs are split into real and imaginary parts to obtain real-valued channels.

\textbf{DyCA} finds modes whose amplitudes best satisfy an ODE system with $m$ linear and $n\!-\!m$ nonlinear equations, thereby separating deterministic dynamics from stochastic noise~\cite{Uhl2024}. We use $m\!=\!2$ for $n\!=\!4$.

\textbf{AVD} takes a different approach: rather than a spatial decomposition, it computes a univariate measure of cross-channel dispersion heterogeneity~\cite{Mallinger2026}. For a dispersion estimate $\delta_{j}(t,w)$ (median absolute deviation, MAD, or standard deviation, SD) per channel over sliding window length $w$, the AVD signal averages squared differences across all channel pairs $(j, j')$ with $1\le j<j'\le N$:
\begin{equation}
\mathrm{AVD}(t,w) = \frac{1}{N(N\!-\!1)} \sum_{1\le j<j'\le N} \bigl(\delta_{j}(t,w) - \delta_{j'}(t,w)\bigr)^{2}.
\end{equation}
Combining two dispersion estimators (MAD, SD) with two window lengths ($w\!\in\!\{10,25\}$ samples) yields four AVD signals, which are stacked as the 4-channel classifier input.

\subsection{Classifier and Training}

Each 4-channel, 2-second window ($4\!\times\!512$ samples at 256\,Hz) is classified as seizure or non-seizure by a deep neural network. To ensure a fair comparison across dimension reduction methods, we conduct an automated architecture and hyperparameter search for each representation (PCA, DyCA, DMD, AVD). The search, adapted from the \emph{autoresearch} framework~\cite{Karpathy2025Autoresearch}, uses a large-language-model (LLM) coding agent. The agent iteratively proposes classifier architectures, loss functions, and training hyperparameters, trains each candidate on a high-performance computing (HPC) cluster, and retains only those experiments that improve the validation area under the receiver operating characteristic curve (ROC-AUC). This yields a separately optimized classifier per method, so that performance differences reflect the input representation rather than classifier tuning.

All models are trained initially with the AdamW optimizer (learning rate $10^{-3}$) and a batch size of 256. To address class imbalance, we use a focal loss with focusing parameter $\gamma=2.0$ and class weights $\alpha=[1,\,10]$ (non-seizure, seizure). Training is stopped early if the validation ROC-AUC does not improve for 25 consecutive epochs. The autoresearcher may adjust all hyperparameters based on validation performance.

Our github repository\footnote{\url{https://github.com/HS-Ansbach-CCS/bmt2026-eeg-seizure-detection-autoresearch}} contains the search code with the \emph{autoresearch} pipeline, the baseline model, and the optimized model architectures.

\section{Results and Discussion}
\subsection{Autonomous Search Outcomes}

\begin{table}[!t]
\centering
\caption{Best-performing classifier architecture per dimension reduction method, as selected by the autonomous search.\label{tab:architectures}}
\begin{tabular}{lp{0.80\columnwidth}}
\toprule
Method & Architecture \\
\midrule
AVD  & MultiTimescale Filterbank (multi-scale spectral features at different temporal resolutions) \\
PCA  & MultiBand CrossCoupling (3-stream architecture; spectral features, bilinear coupling, autocorrelation) \\
DMD  & MultiScale Lyapunov Net (Lyapunov-like trajectory divergence CNN) \\
DyCA & SpectralTemporal DualStream (time-domain and frequency-domain CNN-GRU) \\
\bottomrule
\end{tabular}
\end{table}

Table~\ref{tab:architectures} lists the best classifier architecture selected for each dimension reduction method. The optimal architecture differs across representations. For the variance-based methods AVD and PCA, the search converges on architectures that decompose the input into multiple frequency bands or temporal scales, thereby extracting rich spectral structure from the high-variance components preserved by these representations. The AVD classifier employs a parallel filterbank with four kernel sizes followed by three gated recurrent unit (GRU) streams operating at different temporal resolutions, while the PCA classifier uses learned bandpass filters with bilinear cross-band coupling to capture inter-frequency interactions.

In contrast, the dynamics-based methods DMD and DyCA yield dual-stream architectures that process complementary signal views in parallel (DMD: raw signal plus trajectory-divergence features; DyCA: separate time- and frequency-domain streams). This suggests their discriminative content is distributed across complementary domains, requiring the classifier to fuse heterogeneous features rather than refine a single rich representation.

These architectural differences confirm that representation and classifier cannot be optimized in isolation: pairing all four representations with a single fixed architecture would obscure performance differences attributable to the representation itself.

\subsection{Seizure Detection Performance}

Table~\ref{tab:results} summarizes performance (ROC-AUC and PR-AUC) on both data splits. On the validation set, the variance-based methods AVD (89.78\,\% ROC-AUC) and PCA (89.41\,\%) outperform the dynamics-based methods DMD (84.04\,\%) and DyCA (82.19\,\%) by 5-8\,\% in ROC-AUC; on the held-out test set this gap widens to 11-14\,\%, with AVD maintaining the highest score (88.28\,\%) and DMD/DyCA dropping to 74.56\,\% and 74.85\,\%, respectively. The same ordering holds for PR-AUC on the validation set (AVD 82.58\,\%, PCA 78.57\,\%, DMD 77.39\,\%, DyCA 73.19\,\%); on the test set AVD again leads with PR-AUC of 31.91\,\%.
The much lower absolute PR-AUC on the test split reflects its low seizure prevalence (5.4\,\% (test) and 33.7\,\% (val)), which sets the PR-AUC chance level and precludes comparing absolute PR-AUC across splits.

A plausible explanation for AVD's superiority lies in its multi-scale design: its four configurations span two dispersion estimators (MAD, SD) and two window lengths, so short windows capture transient spikes while longer windows track slower baseline drift, jointly encoding fast and slow seizure-related variance. DyCA's richer, nonlinear modeling assumptions did not translate to better performance, possibly due to constraint rigidity or noise sensitivity in short windows.

\begin{table}[!t]
\centering
\caption{Validation and test performance per dimension reduction method, for the best model and hyperparameters selected by the autonomous search.\label{tab:results}}
\begin{tabular}{lcccc}
\toprule
 & \multicolumn{2}{c}{ROC-AUC (\%)} & \multicolumn{2}{c}{{PR-AUC (\%)}} \\
\cmidrule(lr){2-3}\cmidrule(lr){4-5}
Method & Val & Test & {Val} & {Test} \\
\midrule
AVD  & \textbf{89.78} & \textbf{88.28} & {\textbf{82.58}} & {\textbf{31.91}} \\
PCA  & 89.41 & 85.98 & {78.57} & {25.55} \\
DMD  & 84.04 & 74.56 & {77.39} & {19.30} \\
DyCA & 82.19 & 74.85 & {73.19} & {21.31} \\
\bottomrule
\end{tabular}
\end{table}

This trend is driven by a pronounced generalization deficit in dynamics-based approaches: DMD and DyCA exhibit ROC-AUC drops of 9.48\,\% and 7.34\,\%, whereas AVD and PCA are more stable, with decreases of only 1.50\,\% and 3.43\,\%. As a compact, univariate dispersion measure, AVD appears less prone to patient-specific overfitting than the higher-dimensional modal decompositions of DMD and DyCA. Cross-channel variance heterogeneity thus emerges as a robust seizure signature that transfers better across subjects.
\section{Conclusion}

We systematically compared four dimension reduction methods for EEG seizure detection, each paired with an independently optimized deep learning classifier via an autonomous AI-driven research framework. Variance-based representations (AVD, PCA) consistently outperform dynamics-based decompositions (DMD, DyCA) on both splits, with the gap widening on the test data and AVD achieving the highest test ROC-AUC of 88.28\,\% and the highest test PR-AUC of 31.91\,\%. This indicates that preserving variance structure might be more important than preserving temporal dynamics for seizure detection, and that cross-channel dispersion heterogeneity, as captured by AVD, yields a seizure signature that transfers more reliably across subjects. The autonomous search further reveals that the optimal classifier architecture differs across representations, underscoring the need for joint optimization of representation and classifier in EEG processing pipelines. Our comparison is limited to a single corpus (TUSZ) with window-level binary labels, and cross-subject generalization is assessed only via the predefined test split. Despite these encouraging results, there is further space for improvement. Future work will investigate whether combining variance-based and dynamics-based representations in a complementary fashion can yield additional gains, and whether AVD's cross-subject transferability extends to finer-grained seizure typing, broader clinical cohorts, and different recording conditions. A systematic evaluation of the target dimensionality $n$ remains open.

\medskip
\textsf{\textbf{Author Statement}}\\
This work was supported by Bavarian Ministry of Economic Affairs, Regional Development and Energy (StMWi, Funding number: DIK-2307-0007// DIK0536/01). 
Authors state no conflict of interest.
Not applicable; secondary analysis of the publicly available, de-identified TUSZ Corpus.
Ethical approval: Not applicable; no new human-subject data were collected.
We acknowledge the support and HPC resources provided by the Erlangen National HPC Center of the FAU Erlangen-Nürnberg under the BayernKI project DLonEEGData. BayernKI funding is provided by Bavarian state authorities.
\bibliographystyle{unsrtnat}
\bibliography{deepeeg}
\end{document}